\documentclass{aa}
\usepackage{graphicx}
\usepackage{txfonts}
\usepackage[utf8]{inputenc}
\usepackage{amsmath,amssymb}
\usepackage{lipsum}


\begin{document}

\title{Perturbed precessing ellipses as the building blocks of spiral arms in a barred galaxy with two pattern speeds}
\titlerunning{Perturbed precessing ellipses as the building blocks of spiral arms}
\authorrunning{Harsoula et al.}

\author{M. Harsoula\inst{1}
C.Efthymiopoulos\inst{2}
G. Contopoulos\inst{1} \and
A. C. Tzemos\inst{1}}

\institute{ $^1$Research Center for Astronomy and Applied Mathematics, Academy of Athens,
Soranou Efessiou 4, 115 27 Athens\\
$^2$Department of Mathematics, Tullio Levi-Civita, University of Padua, Via Trieste, 63, 35121 Padova, Italy
\email{  mharsoul@academyofathens.gr; cefthim@academyofathens.gr;   gcontop@academyofathens.gr }}

\date{Received ; accepted}
\maketitle

\begin{abstract}

Observations and simulations of barred spiral galaxies have shown that, in general, the spiral arms rotate at a different pattern speed to that of the bar. The main conclusion from the bibliography is that the bar rotates faster than the spiral arms with a double or even a triple value of angular velocity.  The theory that prevails in explaining the formation of the spiral arms in the case of a barred spiral galaxy 
with two pattern speeds is the manifold theory, where the orbits that support the spiral density wave are chaotic, and are related to the manifolds emanating from the Lagrangian points $L_1$ and $L_2$ at the end of the bar. In the present study, we consider an alternative scenario in the case where the bar rotates fast enough in comparison with the spiral arms and the bar potential can be considered as a perturbation of the spiral potential. In this case, the stable elliptical orbits that  support the spiral density wave (in the case of grand design galaxies) are transformed into  quasiperiodic orbits (or 2D tori) with a certain thickness. The superposition of these perturbed preccesing ellipses for all the energy levels of the Hamiltonian creates a slightly perturbed symmetrical spiral density wave.
\end{abstract}

\section{Introduction}

There are two main theories that prevail nowadays concerning the building blocks of the spiral arms in galaxies.
 In the case of grand design galaxies, the "density wave" theory remains a valid dynamical model with which the spiral structure of many disk galaxies can be described. The density wave theory fits the description of spiral arms better when the spiral amplitude does not exceed a value of 10\%-20\% over a  few pattern rotations. This theory was first developed by Lindblad (1940, 1961), and it was extended by \cite{Lin1964,Lin1966}. \cite {Lind1955} pioneered the orbital description of spiral density waves.  In the density wave theory, the periodic orbits are close to precessing ellipses that support the shape of the spiral structure. Many papers have studied these periodic orbits, in models of grand design galaxies. It has been shown that these orbits can collaborate so that the imposed model matches
with the response model of spiral arms (see   \cite{Cont1970,Cont1971,Cont1975}, \cite{Ber1977}, \cite{Mon1978},
\cite{Cont1986}, \cite{Pat1991}, and \cite{Har2021}).

On the other hand, in the case of barred spiral galaxies, where the perturbation of the bar component is large enough and introduces chaos near corotation,  the prevailing theory for the spiral arms is the "manifold theory" 
(\cite{dan1965}, \cite{vogetal2006}, \cite{rometal2006,rometal2007},  \cite{tsouetal2008, tsouetal2009}, \cite{athetal2009a,athetal2009b,athetal2012}, \cite{haretal2016}). This theory predicts bisymmetric spirals emanating from the ends of galactic bars as a result
of the outflow of matter connected with the unstable dynamics
around the bar’s Lagrangian points $L_1$ and $L_2$. In this case the spiral arms are supported by chaotic orbits having initial conditions along the unstable manifolds emanating from $L_1$ and $L_2$.
More recently, \cite{Efth2020} found empirically that the manifold spirals,
which are computed in an N-body simulation by momentarily
“freezing” the potential and making all calculations in a frame
that rotates with the instantaneous pattern speed of the bar,
reproduce the time-varying morphology of the N-body spirals
rather well. In that simulation it was also found that multiple patterns are demonstrably present.
Observations of barred spiral galaxies  have shown that  the existence of two different pattern speeds for the spiral arms and the bar is a very possible scenario (\cite{Moore1995}, \cite{booetal2005}, \cite{meietal2009}, \cite{spewes2012}, \cite{speroo2016}, \cite{Font2019}). 
The same conclusion emerges from simulations  (\cite{selspa1988}, \cite{rausal1999}, \cite{qui2003},  \cite{quietal2011}, \cite{rocetal2013}, \cite{fonetal2014}). A brief review of the different methods used to determine the pattern speeds
of the Galactic bar and spiral arms of the Milky Way was given in \cite{ger2011}.  In most cases the ratio of the two pattern speeds is 1:2 or 1:3, with the spiral arms always having the lowest value.
In \cite{Efth2020} it has been shown that, in a Milky Way-like model  with two different pattern speeds for the bar and the spiral arms, the manifold theory can still be valid if the spiral potential is considered as a perturbation of the bar's potential. As a consequence,  the unstable Lagrangian points $L_1$ and $L_2$ of the pure bar model are continued in the full model as periodic orbits, or as epicyclic “Lissajous-like” unstable orbits.\\
In the present paper we reverse the idea of \cite{Efth2020} and we consider the potential of the bar as a perturbation of the spiral potential.
An important quantity in this connection is  the  $Q$-strength, which gives an  estimate of the relative importance of the 
bar's and the spirals' non-axisymmetric force perturbations. The $Q$-strength at 
fixed radius $r$ (e.g., \cite{butetal2009}) is defined for the bar as
\begin{equation}\label{qstr}
Q_b(r)={F_{b,t}^{max}(r)\over \langle F(r)\rangle}, 
\end{equation}
where $F_{b,t}^{max}(r)$ is the maximum tangential  
force generated by the potential term $V_{bar}$ at the distance $r$, with respect to all azimuths $\phi$. Moreover, $\langle F(r) \rangle$  
is the average radial force at the same distance,  with respect to $\phi$, generated 
by the potential $V_{ax}+V_{bar}+V_{sp}$.  The  $Q$-strength of the bar is always much larger than the one of the spiral arms (\cite{butetal2005,butetal2009}, \cite{Durbala2009} ). However, the  $Q$-strength alone is not sufficient to state the importance of the bar’s perturbation in relation to the spiral perturbation. In the case where the bar rotates much faster that the spiral arms, an observer rotating with the spiral arms would see the bar as a kind of a bulge (an average axisymmetric component). In such a case,  the  potential of the bar can be considered as a perturbation of the potential of the spiral arms. Therefore, in such a case, a generalization of the density wave theory can be considered.  
Under this assumption, we construct a semianalytical algorithm, with the help of a normal form algorithm, with the aim of eliminating time dependence from the Hamiltonian (due to the difference in the pattern speeds). Then, in this new Hamiltonian, the precessing ellipses that support the spiral structure in the case of grand design galaxies is transformed into  quasiperiodic orbits forming ellipses with a certain thickness. When we superimpose all of them at different energy levels, we can again extract a spiral density wave by calculating the isodensities, using an image processing method described below (Sect. 4).

This paper contains the following: In Sect. 2 we give a description of the galactic model that we use in our study.   In Sect. 3 we quote the normal form construction of the Hamiltonian corresponding to the galactic model and describe the way of locating the stable periodic orbits that are responsible for the spiral density way. We also plot these orbits in the old variables before the normal form transformation in order to see how the precessing ellipses have been deformed  and if we can still detect a spiral density wave. In Sect. 4 we make a parametric study in order to see how the mass and the pattern speed of the bar affects the outcome of the spiral density wave derived from the procedure described in Sect. 3. Finally, in Sect. 5 we summarize the conclusions of our study.

\section{The model}

We consider a model of a Milky Way-like spiral galaxy that contains a combination of a bar, an axisymmetric, and a spiral potential used in \cite{Pett2014}:
\begin{equation}
V =V_{ax}+V_{sp}+V_{bar}.
\end{equation}
The axisymmetric potential $V_{ax}$ is composed of a disk,  a halo, and the axisymmetric part of the bar's potential, which plays the role of a central bulge:
\begin{equation}\label{vax}
V_{ax}=V_{d}(r)+V_{h}(r)+V_{b_0}(r).
\end{equation}
For the disk potential $V_{d}$ , we use a Miyamoto-Nagai model (Miyamoto
and Nagai, 1975) given by the relation 
\begin{equation}
V_{\rm{d}}=\frac{-G M_{\rm{d}}} { \sqrt{ r^{2}+ \left(a_{\rm{d}}+\sqrt{z^2+b_{\rm{d}}^2} \right)^{2}} }
,\end{equation}
where $M_{\rm{d}}=8.56 \times 10^{10}$ $M_\odot$ is the total mass of the disk, $a_{\rm{d}}=5.3$ kpc, and $b_{\rm{d}}=0.25$ kpc. In order to have a 2D disk model,  we take $z=0$ and $r=\sqrt{x^2+y^2}$.\\
The halo
potential is a $\gamma$-model (\cite{Den1993}) with parameters as in
Pettitt et al. (2014),
\begin{equation}\label{pothalo}
V_{\rm{h}}=\frac{-GM_{\rm{h(r)}}}{r}-\frac{-GM_{\rm{h,0}}}{\gamma r_{\rm{h}}}
\left[-\frac{\gamma}{1+(r/r_{\rm{h}})^\gamma}+\ln\left(1+\frac{r}{r_{\rm{h}}}\right)^\gamma\right]_r^{r_{h,max}}
,\end{equation}
where $r_{h,max}=100$ kpc, $\gamma=1.02$, $M_{\rm{h,0}}=10.7 \times 10^{10} M_\odot$,
and $M_{\rm{h(r)}}$ is given by the function:
\begin{equation}\label{mhr}
M_{\rm{h(r)}}=\frac{M_{\rm{h,0}}(r/r_{\rm{h}})^{\gamma+1}}{1+(r/r_{\rm{h}})^\gamma}~~.
\end{equation}
The spiral potential is given by the value $V_{\rm{sp}}$ (for $z=0$) of the 3D logarithmic spiral model $V_{\rm{sp}}(r,\phi,z)$ introduced by Cox and Gomez (2002) (see Formula (19) in Efthymiopoulos et al. 2020): 
\begin{equation}
V_{\rm{sp}}= 4 \pi G  h_{\rm{z}} \rho_{0}~ G(r)~\exp \left(- \left(\frac{r- r_{\rm{0}}} { R_{\rm{s}}} \right) \right) {\frac{C}{K B}}~ 
\cos \left[ 2 \left(\varphi-\frac{\ln(r/ r_{0} )}{\tan(\alpha)} \right) \right]
,\end{equation}
where
\begin{equation}
K=\frac{2}{r  |~\sin(\alpha)~|   } , ~~~B= \frac{1+K h_{z}+0.3 (K  h_{z})^{2} }{1+0.3K  h_{z}}
,\end{equation}
and $C=8/(3 \pi)$, $h_z =0.18$ kpc, $r_0=8$ kpc, $R_s=7$ kpc, and $a=-13^{o}$ is the pitch angle of the spiral arms.  The function $G(r)$ plays the role of a smooth envelope determining the radius  
beyond which the spiral arms are important. We adopt the form 
$G(r)= b-c\arctan(R_{s_0}-r)$, with $R_{s_0}=6$~kpc, $b=0.474$, and $c=0.335$. The spiral density is $\rho_{0} = 15 \times10^7$~$M_\odot~\rm{kpc}^{-3}$ in the  model under study. This value of the spiral density is chosen so as to yield a spiral $F$-strength  value of 15\%, consistent with those reported in the literature for the case of an intermediate spiral perturbation (see \cite{Pat1991} and \cite{gros2004} for grand design galaxies, and \cite{{Sar2018}} and \cite{Pett2014} for the Milky Way. The $F$-strength value is given by the ratio of the maximum total force of the spiral perturbation over the radial force of the axisymmetric background:

\begin{equation}\label{qall}
F_{\rm{all}}(r)=\frac{\left\langle F_{\rm{sp}}(r)\right\rangle}{F_{r}(r)}=
\frac{\left\langle\sqrt{\left(\frac{1}{r}\frac{\partial V_{\rm{sp}}}{\partial \varphi}\right)^2 +
\left(\frac{\partial V_{\rm{sp}}}{\partial r}\right)^2}\right\rangle_{\rm{max}}}{\frac{\partial V_{\rm{ax}}}{\partial r}}
\end{equation} 
(see Fig.1 and corresponding text of \cite{Har2021} for a detailed explanation of the role of the $F$-strength value). 

 The bar's potential potential is as in \cite{lonmur1992}: 
\begin{equation}\label{potbar}
V_b=\frac{GM_b}{2a}\ln \left(\frac{x-a+T_-}{x+a+T_+}\right),
\end{equation}
with  
\begin{equation}\label{potbart}
T_{\pm} = \sqrt{\left( a \pm x \right)^2+y^2+\left(b+\sqrt{c^2+z^2}\right)^2},
\end{equation}
where $M_b$ is the mass of the bar. In our study we use three different values of $M_b$, namely  $M_b=(6.25,~ 3, ~1) \times 
10^{10} M_\odot$,  where $M_\odot$ is the solar mass, $a=5.25$~kpc, $b=2.1$~kpc, and $c=1.6$~kpc. The values of $a ~and b$ 
set the bar's length along the major and minor axes in the disk plane ($x$ and $y,$ 
respectively), while $c$ sets the bar's thickness in the z-axis (see 
\cite{ger2002}, \cite{ratetal2007}, \cite{caoetal2013}). These values 
are chosen so as to bring the bar's corotation (for $\Omega_{b}=45$~km/s/kpc) 
to the value (specified by the $L_{1,2}$ points distance from the center) 
$R_{L1,2} = 5.4$~kpc. Assuming corotation to be at $1.2 - 1.3$ times the bar length, as seen in the literature, the latter turns to be about $4$Kpc with the adopted parameters. 
We set $z=0$, because we deal with the 2D model, and finally, Eq. (\ref{potbar}) in polar coordinates $(r, \theta)$, in the inertial frame of reference, takes the form
\begin{flalign}\label{polbar}
& V_b(r,\theta)=\frac{GM_b}{2a} \nonumber \\
& \times \ln \left(\frac{r \cos (\theta)-a+ \sqrt{(r \cos (\theta)-a)^2+(b+c)^2+r^2 \sin^2(\theta)}}{r \cos (\theta)+a+ \sqrt{(r \cos (\theta)+a)^2+(b+c)^2+r^2 \sin^2(\theta)}}\right).
\end{flalign}
We now make a Taylor expansion of Eq. (\ref{polbar}) with respect to $\cos (\theta)$ and rewrite the bar's potential as the sum of $\cos(m \theta)$ terms, where $m=0,2,4,6$. Then, Eq. (\ref{polbar}) takes the form
\begin{flalign}\label{cosbar}
& V_b(r,\theta)=V_{b0}(r)+ \sum_{m=2}^{6} V_{bm}(r,\theta)= \nonumber \\
& A_{b0}(r)+A_{b2}(r) \cos (2\theta)+A_{b4}(r) \cos (4\theta)+A_{b6}(r) \cos (6\theta).
\end{flalign}
The bar's  and the spiral's potential,  in the inertial frame of reference and in polar coordinates $(r,\theta),$ are given by the following relations:
\begin{equation}\label{barin}
V_{b_{in}}=V_b(r,\theta-\Omega_b~t),~~~V_{{sp}_{in}}=V_{sp}(r,\theta-\Omega_{sp}~t),
\end{equation} 
where $\Omega_{b}$ is the pattern speed of the bar and $\Omega_{sp}$ is the pattern speed of the spiral arms  in the inertial frame of reference.
The Hamiltonian of the total potential, in the rotating frame of reference of the spiral potential, is then written in the form
\begin{eqnarray}\label{hamin}
\nonumber H_{in}(r,f,f_2)= \frac{p_r^2}{2}+ \frac{p_f^2}{2r^2}-\Omega_{sp} p_f+ V_{ax}(r)\\
+ \sum_{m=2}^{6}V_{bm}(r,f+f_2)+V_{sp}(r,f)+ \Delta \Omega J_2,
\end{eqnarray}
where $p_r$ is the radial velocity per unit mass, $p_f$ is the angular momentum of the unit mass in the rest frame of reference, $f=\theta-\Omega_{sp}~t$, $f_2=\Delta \Omega ~t =(\Omega_{sp}-\Omega_{b})~ t$, and $J_2$ is the canonically conjugate action of the angle $f_2$. The axisymmetric part of the potential $V_{ax}(r),$ as given by Eq. (\ref{vax}), includes the axisymmetric part of the bar's potential $V_{b0}(r)$, while the non-axisymmetric part of the bar's potential  $V_{bm}(r,f+f_2)$ includes, in a first approximation, only the $m=2$ terms of the bar's potential,  in order to facilitate the construction of the normal form:
\begin{equation}\label{bar2}
V_{b_2}(r,f+f_2)=A_{b_2}(r) \cos \left (2(f+f_2) \right).
\end{equation}

\section{Normal form construction}

 We now make a Taylor expansion of the bar's potential  $V_{\rm{b_2}}(r,f,f_2)$ around the radius of the circular orbit $r_c$ and the corresponding angular momentum $p_c$, up to the second order, making the following replacements:
\begin{equation}\label{dr}
~~~~~~~~~~~~~r=r_c+ \delta r,
~~~~~~p_f= p_c + J_u,
\end{equation}
 where $\delta r$ is a small perturbation of the radius in relation to the radius of the circular orbit $r_c$ and $J_u$ is a small perturbation of the angular momentum in relation to the angular momentum of the circular orbit $p_c$.
Finally, we make a canonical transformation in action-angle variables $(\delta r,p_r)\rightarrow(J_r,f_r)$: 
\begin{equation}\label{Jr}
~~~~~~~~~~~~~\delta r=\sqrt{2 J_r/\kappa _c} \sin{f_r},
~~~~~~p_r=\sqrt{2 J_r \kappa_c} \cos{f_r},
\end{equation}
where $k_c(r_c)$ is the epicyclic frequency at $r=r_c$, given by the relation
\begin{equation}\label{kc}
\kappa _c=\sqrt{\frac{d^2V_{ax}(r_c)}{d r^2_c}+\frac{3}{r_c}\frac{dV_{ax}(r_c)}{dr_c}},
\end{equation}
and $p_c$ is the angular momentum of a star moving in a circular orbit with radius $r_c$.  Moreover, $p_c$ is related to the angular velocity $\Omega(r_c)$ of the star with the relation $p_c=\Omega(r_c) r_c^2$. The angular velocity $\Omega(r_c)$, under the action of the axisymmetric potential only, is given by the relation
 \begin{equation}\label{omc}
\Omega_c=\sqrt{\frac{1}{r_c} \frac{dV_{ax}(r_c)}{d r_c}}.
\end{equation} 
Then, the Hamiltonian (\ref{hamin}) takes the following form:
\begin{flalign}\label{newHam}
& H_{new}(J_u,J_r,J_2,f,f_r,f_2)= H_c+ \Delta \Omega J_2+ \kappa_c J_r  +(\Omega(r_c)-\Omega_{sp})Ju  \nonumber \\
&  
+   \sum_{m_1,m_2,m_3} a_{m_1,m_2,m_3} (J_r)  _{\sin}^{\cos}(m_1 f+m_2 f_2+m_3 f_r),
\end{flalign}
where  $m_1,m_2, m_3$ can take the values $0,\pm 1,\pm 2$.  The notation $_{\sin}^{\cos}(m_1 f+m_2 f_2+m_3 f_r)$ in the summation of Eq. (\ref{newHam}) means that there are both terms with $  \sin (m_1 f+m_2 f_2+m_3 f_r) $ and $   \cos (m_1 f+m_2 f_2+m_3 f_r) $. $\Delta \Omega= \Omega_{sp}-\Omega_{b}$ and $H_c$ is a constant term that can be omitted.

We now want to construct a normal form of the Hamiltonian (\ref{newHam}) in order to eliminate the terms that contain the angle $f_2= \Delta \Omega~ t$ and make the Hamiltonian nonautonomous (for a tutorial on the normal form construction, see \cite{eft2012}). We introduce a formal notation to account for this consideration: in front of every term
in Eq. (\ref{newHam}), we introduce a factor, $\lambda^s$ , hereafter called the ‘‘book-keeping
parameter’’, which is a constant with a numerical value equal to $\lambda$=1, while $s$ is a positive integer exponent whose value, for every term in  (\ref{newHam}), is selected so as to reflect our consideration regarding the order of smallness we estimate a term
to be of in the Hamiltonian. Thus, considering the leading terms $\Delta \Omega J_2+ \kappa_c J_r +(\Omega(r_c)-\Omega_{sp})Ju$ to be of order zero, we put $\lambda^0$ in front of them. On the other hand, considering the remaining terms to be of a similar order of smallness (i.e., first order), we
put a factor, $\lambda^1$ , in front of them. Then Eq. (\ref{newHam}) becomes

\begin{flalign}\label{hnewl}
& H_{new}(J_u,J_r,J_2,f,f_r,f_2)= \lambda^0 \left(\Delta \Omega J_2+ \kappa_c J_r +(\Omega(r_c)-\Omega_{sp})Ju \right) +\nonumber \\
& \lambda ^1 \left(\sum_{m_1,m_2,m_3} a_{m_1,m_2,m_3} (J_r)  _{\sin}^{\cos}(m_1 f+m_2 f_2+m_3 f_r)\right) = \\
& \lambda^0 H_0 + \lambda^1 H_1 \nonumber,
\end{flalign}
 where $H_0$ is the integrable part of the Hamiltonian depending only on actions, and  $H_1$ is the part depending on action $J_r$ as well as on all the angles $f$, $f_2$, and $f_r$. 

In order to construct the normal form of the Hamiltonian  (\ref{hnewl}) and eliminate the dependence on $f_2$, we find the corresponding generating  function $\chi$ using the homological equation
\begin{equation}\label{homol}
\lbrace H_0, \chi \rbrace + h_{kill}=0,
\end{equation} 
where $\lbrace ~ \rbrace$ defines the Poisson bracket and 
\begin{equation}
H_0=\Delta \Omega J_2+ \kappa_c J_r +(\Omega(r_c)-\Omega_{sp})Ju 
\end{equation}
is the integrable part of the Hamiltonian (\ref{newHam}) containing the three basic frequencies $\Delta \Omega$, $\kappa_c$, and $\Omega(r_c)-\Omega_{sp}$. The term  $h_{kill}$ is the part of the Hamiltonian (\ref{newHam}) that contains the angle $f_2$ that we want to eliminate:
\begin{equation}\label{hkill}
h_{kill}= \sum a_{m_1,m_2,m_3} (J_r) ~~ _{\sin}^{\cos}(m_1 f+m_2 f_2+m_3 f_r)~~~ with ~~~m_2 \neq 0.
\end{equation} 
Solving Eq. (\ref{homol}) we find the generating function $\chi$, which is of the form
\begin{flalign}\label{gener}
& \chi = \sum a_{m_1,m_2,m_3} (J_r) ~~\frac{ _{-\cos}^{~~\sin}(m_1 f+m_2 f_2+m_3 f_r)}{(m_1(\Omega(r_c)-\Omega_{sp})+m_2(\Omega_{sp}-\Omega_b)+m_3 \kappa_c)} \nonumber \\
& m_2 \neq 0.
\end{flalign} 
The first-order normal form of the Hamiltonian (\ref{newHam}) is found by the relation
\begin{equation}\label{hnorm}
 H_{norm}^{(1)}=H_{new} + \lbrace H_0, \chi \rbrace =Z_0(J_u,J_r,J_2)+ \lambda~ Z_{norm}^{(1)}(J_r,f,f_r),
\end{equation}
where $Z_0=H_0$ and $Z_{norm}^{(1)}(J_r,f,f_r)=\sum a'_{m_1,m_3} (J_r) ~~ _{\sin}^{\cos}(m_1 f+m_3 f_r) $.\\
 The second-order normal form of the Hamiltonian (\ref{newHam})  is found by the relation
\begin{flalign}\label{hnorm2}
&  H_{norm}^{(2)}=H_{new} + \lbrace H_0, \chi \rbrace + \lbrace H_1, \chi \rbrace + \frac{1}{2} \lbrace \lbrace H_0, \chi \rbrace,  \chi \rbrace = \nonumber \\
& Z_0(J_u,J_r,J_2)+\lambda Z_{norm}^{(2)}(J_r,f,f_r) +O(\lambda^2),
\end{flalign}
where $Z_0=H_0$,~ $Z_{norm}^{(2)}(J_r,f,f_r)=\sum b'_{m_1,m_3} (J_r) ~~ _{\sin}^{\cos}(m_1 f+m_3 f_r), $ and $O(\lambda^2)=\sum b'_{m_1,m_2,m_3} (J_r) ~~ _{\sin}^{\cos}(m_1 f+m_2 f_2 +m_3 f_r)$.
The Hamiltonian (\ref{hnorm}) and the zero- and first-order terms of the Hamiltonian  (\ref{hnorm2}) (in the book keeping parameter $\lambda$) are  now transformed in the new variables $(r_{new},f_{new},p_{r_{new}},p_{f_{new}})$ using Eqs. (\ref{dr}) and  (\ref{Jr}) :
\begin{flalign}\label{hnew}
&  H_{norm}^{(1)}(r_{new},f_{new},p_{r_{new}},p_{f_{new}})=Z_0+ \lambda~ Z_{norm}^{(1)}(r_{new},f_{new},p_{r_{new}},p_{f_{new}}), \nonumber \\
& H_{norm}^{(2)}(r_{new},f_{new},p_{r_{new}},p_{f_{new}})=Z_0+ \lambda~ Z_{norm}^{(2)}(r_{new},f_{new},p_{r_{new}},p_{f_{new}})\nonumber\\
& +O(\lambda^2).
\end{flalign}
We find the main families of stable and unstable periodic orbits by the  Hamiltonian $H_{norm}^{1}$ of Eq. (\ref{hnew}) using the method described in \cite{Har2021}.
The main family of stable periodic orbits that has the form of precessing ellipses and supports the spiral density wave is named the  $x_1$ orbit after \cite{Cont1975}, who introduced the nomenclature of the families of periodic orbits in a  spiral galaxy.   The other two families of periodic orbits are named $x_2$ (stable periodic orbits) and $x_3$, (unstable periodic orbit) and do not support the spiral density waves.

\subsection{Finding the $x_1$ stable periodic orbits}

The stable periodic orbits of the  $x_1$ family correspond to the continuation of the circular orbits of the axisymmetric part of the potential $V_{ax}$ (Eq. \ref{vax}) into the region of the  2:1 resonance (inner Lindblad resonance).

The method of finding these orbits is thus: we fix a radius of a circular orbit $r_c$ and for this specific radius, we calculate the first order normal form $H_{norm}^{(1)}(r,f,p_r,p_f)$  (we eliminate the subscript "new" for brevity) of Eq. (\ref{hnew}) following the procedure described in the previous section.

Then, we find the $x_1$ stable periodic orbit, corresponding to this specific radius $r_c$, by the use of a stroboscopic Poincar\'{e} section ($r$, $p_r$) for $f=2 \kappa \pi$. The Poincar\'{e} section of Fig. \ref{Fig1}a corresponds to the radius $r_c$=7kpc and the central point ($r_{x_1},~p_{r{x_1}}$) corresponds to the stable periodic orbit of the $x_1$ family.  The curves around the stable periodic orbit $x_1$ correspond to quasiperiodic orbits.   Using a Newton Rapshon iterative method, we locate the periodic orbit $x_1$, selecting as initial condition the center of the island of stability ($r_{x_1}$, $f_{x_1}$, $p_{r_{x_1}}$, $p_{f_{x_1}}$).  Then we integrate these initial conditions using a Runge Kutta method of integration and the Hamilton equations of motion:
\begin{figure}
\centering
\includegraphics[scale=0.45]{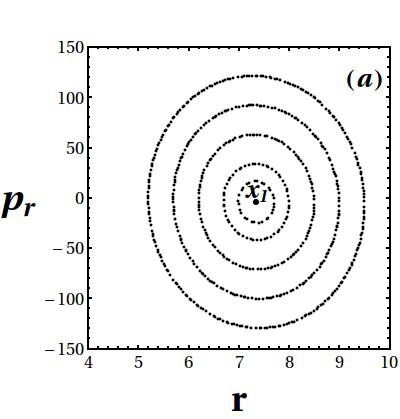}
\includegraphics[scale=0.34]{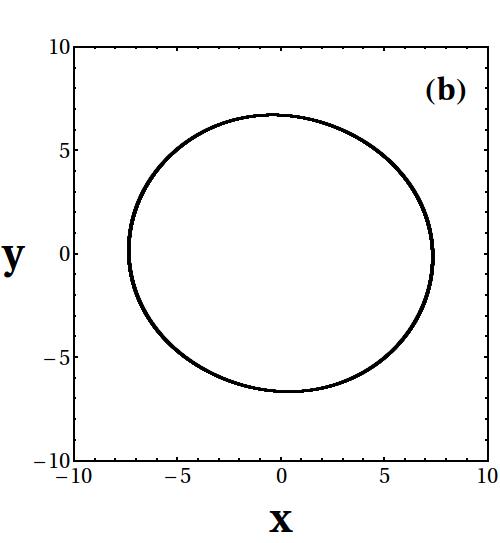}
\caption{ (a) Stroboscopic Poincar\'{e} section ($r$, $p_r$) for $f=2 \kappa \pi$ and $r_c=7.0$, $M_b=6.25 \times 10^{10} M_\odot$, $\Omega_b=45$~km/s/kpc, and $\Omega_{sp}=15$~km/s/kpc. (b) The stable periodic orbit of the $x_1$ family.} 
\label{Fig1}
\end{figure}

\begin{figure}
\centering
\includegraphics[scale=0.35]{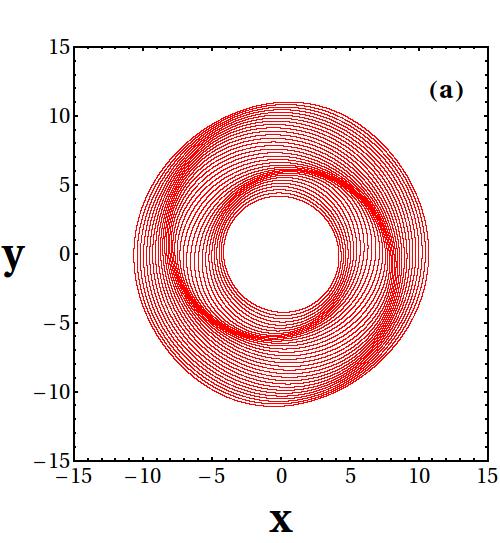}
\includegraphics[scale=0.35]{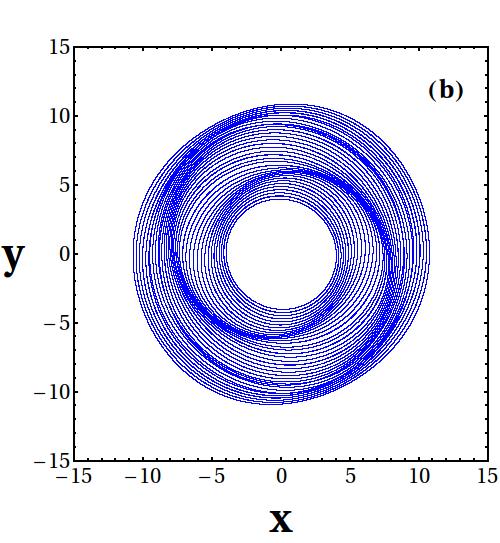}
\caption{(a) The spiral density wave derived from the periodic orbits of the $x_1$ family for the normal form Hamiltonian (\ref{hnew}) of first order $H_{norm}^{(1)}$ and (b) of second order $H_{norm}^{(2)}$, for $M_b=6.25 \times 
10^{10} M_\odot$ (a strong bar), $\Omega_b=45$~km/s/kpc, and $\Omega_{sp}=15$~km/s/kpc.} 
\label{Fig2}
\end{figure}

\begin{figure}
\centering
\includegraphics[scale=0.35]{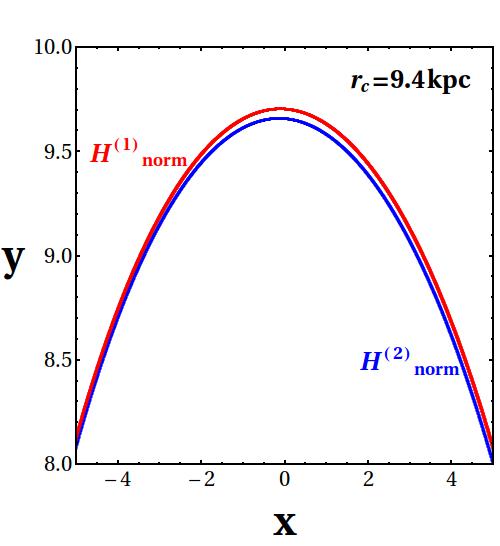}
\caption{Difference between the periodic orbits of the $x_1$ families derived by the first-order normal form $H_{norm}^{(1)}$ (red curve) and the second-order normal form $H_{norm}^{(2)}$ (blue curve) for $r_c=9.4$kpc, where the greater deviation between the two orbits occurs.  } 
\label{Figdiaf}
\end{figure}

\begin{flalign}
& \frac{dr}{dt}=\frac{\vartheta H_{norm}^{(1)}}{\vartheta p_r},~~~~\frac{df}{dt}=\frac{\vartheta H_{norm}^{(1)}}{\vartheta p_{f}}, \nonumber \\
& \frac{dp_r}{dt}=-\frac{\vartheta H_{norm}^{(1)}}{\vartheta r},~~~~\frac{dp_{f}}{dt}=-\frac{\vartheta H_{norm}^{(1)}}{\vartheta f}.
\end{flalign}
The orbit derived corresponds to the $x_1$ family and has an approximately elliptic shape (Fig. \ref{Fig1}b ). 
We repeat this procedure for all the radii $r_c$ between 4.4kpc and 10 kpc with a length step $dr=0.2$kpc, and superimpose the elliptical orbits in Fig. \ref{Fig2}(a). We observe that these precessing ellipses form a spiral density wave located  between their apocenters and pericenters.

If we repeat the procedure described above for the second-order normal form $H_{norm}^{(2)}$ of Eq. (\ref{hnew}), we derive a similar spiral density wave with almost no differences from the one derived from the first-order normal form (see Fig. \ref{Fig2}(b)). 
In fact, the orbits corresponding to the same radius $r_c$ derived by the first- and the second-order normal forms almost coincide, except for a small range of radii around $r_c=9.4$kpc and up to $r_c=9.6$kpc, where there is a small deviation between them.  We see that this small deviation results in the appearance of some weak secondary spiral structures, seen as bifurcations from the main spiral, which is something that one can observe in real galaxies.
In Fig. \ref{Figdiaf} we plot a part of the orbits corresponding to the radius $r_c=9.4$kpc derived from the first-order normal form $H_{norm}^{(1)}$ (in red) and from the second order normal form $H_{norm}^{(2)}$ (in blue). For this specific radius, we observe the maximum deviation between the two cases.

In general, the spiral density wave derived from the precessing ellipses of the $x_1$ family, in both cases, is  well defined.
So, we can now continue with the transformation of the coordinates of the elliptical orbits of the $x_1$ family to the old variables $(r_{old},f_{old},p_{r_{old}},p_{f_{old}})$ (before the normal form construction) using the orbits derived from the first-order normal form Hamiltonian $H_{norm}^{(1)}$, in order to see how these ellipses are deformed.

\subsection{Transformation to the old variables }

A back transformation to the old variables $(r_{old},f_{old},p_{r_{old}},p_{f_{old}})$ corresponding to the initial,  time-dependent Hamiltonian  (\ref{hamin}), is necessary in order to see how the precessing ellipses forming a well-defined spiral density wave are deformed (from the first- and the second-order normal form constructions) for $\Omega_b=45$~km/s/kpc and $\Omega_{sp}=15$~km/s/kpc. 

We make the back transformation of the spiral density wave derived from the  first-order normal form $H_{norm}^{(1)}$  in the old variables, that correspond to the Hamiltonian  (\ref{hamin}), which includes the time-dependent angle $f_2= \Delta \Omega ~t$. This transformation is made using the generating function $\chi$.
Using the new variables of each periodic orbit $x_1$ in polar coordinates, $(r_{new},f_{new},p_{r_{new}},p_{f_{new}}),$ and for a specific radius of the circular orbit $r_c$, we find the old variables 
$(r_{old},f_{old},p_{r_{old}},p_{f_{old}})$ using the following relations:
\begin{flalign}\label{rold}
&\nonumber r_{old}=r_{new}+\lbrace r_{newp}, \chi \rbrace~~~~~~with~~r_{newp}=r_c+ \sqrt{2 J_r/\kappa_c} \sin{f_{r_{new}}},\\
&\nonumber f_{old}=f_{new}+\lbrace f_{newp}, \chi \rbrace~~~~~~with~~f_{newp}=f_{new},\\
& p_{r_{old}}=p_{r_{new}}+\lbrace p_{r_{newp}}, \chi \rbrace~~~~~~with~~p_{r_{newp}}= \sqrt{2 J_r~ \kappa_c} \cos{f_{r_{new}}},\\
& \nonumber p_{f_{old}}=p_{f_{new}}+\lbrace p_{f_{newp}}, \chi \rbrace~~~~~~with~~p_{f_{newp}}= p_c +J_u,
\end{flalign}
where $\chi$ is the solution of Eq. (\ref{homol}), $p_c=\Omega(r_c)~r_c^2$, and $\lbrace ~ \rbrace$ defines the Poisson bracket. Using Eq. (\ref{Jr}), we can replace $\cos{f_{r_{new}}},~~\sin{f_{r_{new}}}$, and $J_r$ in the relations in Eq. (\ref{rold}) as follows: $\cos{f_{r_{new}}}=p_{r_{new}} / \sqrt{2 J_r~ \kappa_c}$, $\sin{f_{r_{new}}}=(r_{new}-r_c)/\sqrt{2 J_r~ / \kappa_c}$, and  $J_r=(1/(2 \kappa_c))(p_{r_{new}}^2+ \kappa_c (r_{new}-r_c)^2)$.
Using Eq. (\ref{rold}), we find the perturbed precessing ellipse for each radius $r_c$ of the circular orbit, which includes the information of the second pattern speed of the bar $\Omega_b$ in cartesian coordinates ($x_{old}=r_{old}~\cos(f_{old}),~y_{old}=r_{old}~\sin(f_{old})$).

\begin{figure*}
\centering
\includegraphics[scale=0.6]{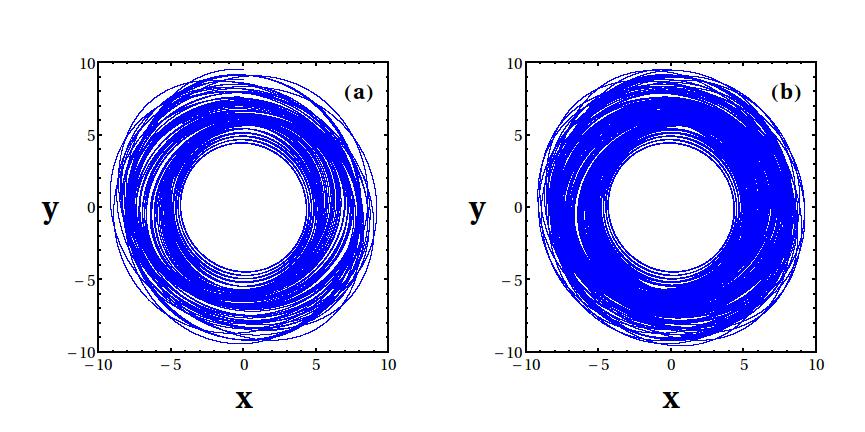}
\caption{Perturbed precessing ellipses of Fig. \ref{Fig2}: (a) for a short integration time corresponding to ten epicyclic periods;  and (b) a longer integration time corresponding to 20 epicyclic periods (see text) for $M_b=6.25 \times 
10^{10} M_\odot$ (a strong bar), $\Omega_b=45$~km/s/kpc, and $\Omega_{sp}=15$~km/s/kpc.} 
\label{Fig3}
\end{figure*}

Figure  \ref{Fig3} shows the perturbed precessing ellipses for the radii  of Fig. \ref{Fig2}, derived using the relations in Eq. (\ref{rold}). In particular, Fig. \ref{Fig3}a consists of orbits integrated for a short time, namely for ten epicyclic periods ($t=10~T_{epi}$), and Fig. \ref{Fig3}b consists of orbits integrated for a longer time, namely for 20 epicyclic periods ($t=20~T_{epi}$), where $T_{epi}=2 \pi / \kappa_c$ and $\kappa_c$ is the epicyclic frequency given by Eq. (\ref{kc}) for each radius $r_c$.

\begin{figure}
\centering
\includegraphics[scale=0.30]{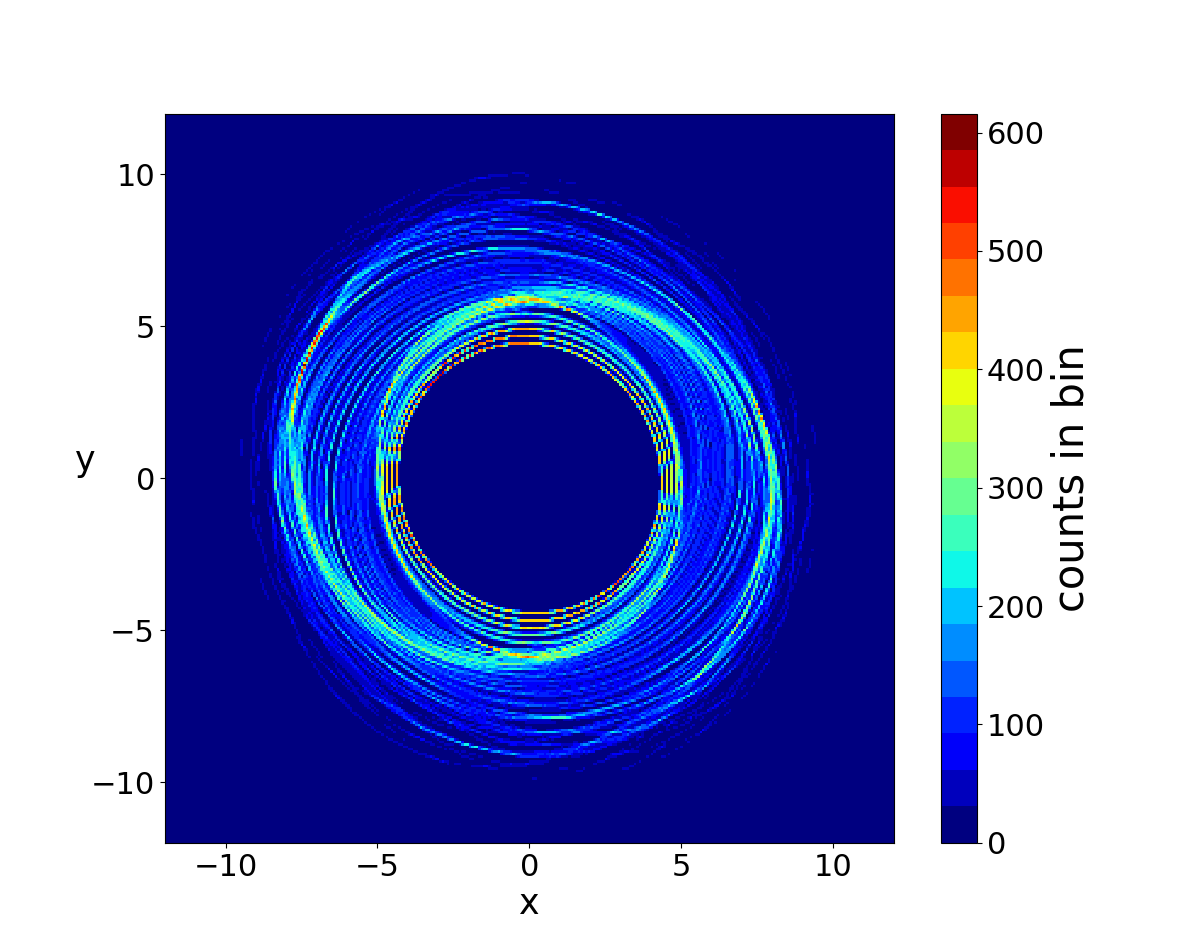}
\caption{Processed image of Fig. \ref{Fig2}b using the method described in the text. A well-defined and intense spiral density wave is presented. } 
\label{Fig4}
\end{figure}

In Fig. \ref{Fig3}a it is quite difficult to distinguish the hidden spiral density wave, while in Fig. \ref{Fig3}b it is impossible to distinguish any pattern at all. For this reason  we used an image processing method in order to recognize the hidden pattern of Figs.  \ref{Fig3}a and 4b.
The method is described as follows: let $G$ be  a  grid of $L~\times L $ square cells (with $L=300$) that covers the dimensions of the galactic model (i.e., $-12 \leq x \leq 12 $kpc and $-12 \leq y \leq 12 $kpc). If $S=[x(t_i),y(t_i)]$ is a time series of a single star trajectory on the configuration space, with $t_i=i\Delta t, i=0, 1, 2, \dots, I$,   collected with a time step $\Delta t=10^{-4}~(2\pi/\kappa_c)$ (where $\kappa_c$ is the epicyclic frequency of each orbit given by Eq. (\ref{kc})) and  $t=t_I$ is the total time of the integration of the orbit, then we define the ``single star trajectory distribution''  $P_s(x_j,y_k; t_i)$ over $G$ around the points $(x_j, y_k)=(j\Delta x, k\Delta_y)$ with $j, k=-N, -N+1,\dots, N-1, N$, where $N=|x_{max}/\Delta x|=|y_{max}/\Delta y|$ and $x_{max}=y_{max}=12$kpc, as: 
\begin{align}
P_S(x_j, y_k; t_i)=C_S(x_j, y_k; t_i), 
\end{align}
 where $C_S$ is the number of points of the sample $S$ inside the square  cell defined by $x_j-\Delta x/2\leq x< x_j+\Delta x/2,~ y_k-\Delta y/2\leq y< y_k+\Delta y/2$. Therefore, $C_s(x_j, y_k; t_i)$ is the ``single star trajectory occupation number'' of the cell $(x_j, y_k)$
 from $t=0$ up to $t=t_I$.

The above considerations are easily extended in the case of an ensemble of $M$ trajectories evolved up to $t=t_I$.

In Fig. \ref{Fig4} we see the result of the processing of Fig. \ref{Fig3}b with the method described above. The  hidden spiral structure is revealed to be well defined and intense, but slightly deformed in relation to the one of Fig. \ref{Fig2} .

\section{Parametric study} 

The shape of the hidden density wave and how well the spiral structure is maintained in a galactic model containing two different pattern speeds, depend on several parameters, such as the mass of the bar $M_b$ in Eq. (\ref{potbar}) and the pattern speed of the bar $\Omega_b$. The perturbation of the precessing ellipses of Fig. \ref{Fig2} depends on the generating function given by Eq. (\ref{gener}). While the mass of the  bar $M_b$ affects the coefficients  $ a_{m_1,m_2,m_3}$ of Eq. (\ref{gener}), the  pattern speed of the bar $\Omega_b$ is found in the denominator of Eq. (\ref{gener}) in the term $\Delta \Omega = \Omega_{sp}-\Omega_b$. Therefore, the perturbation of the precessing ellipses becomes greater for greater values of $M_b$ and for smaller absolute values of $\Delta \Omega$=$\Omega_{sp}-\Omega_b$, in other words smaller values of $\Omega_b$.

We now make a parametric study for different values of $M_b$ and $\Delta \Omega$ in order to test the limits inside which the spiral density wave is still detectable.  

\subsection{The role of the mass of the bar}

In order to test the effect of the mass of the bar, we use two smaller values of the value $M_b$, namely $M_b=3 \times 
10^{10} M_\odot$ and $M_b=1 \times 
10^{10} M_\odot$, and repeat the procedure described above. 
\begin{figure*}
\centering
\includegraphics[scale=0.30]{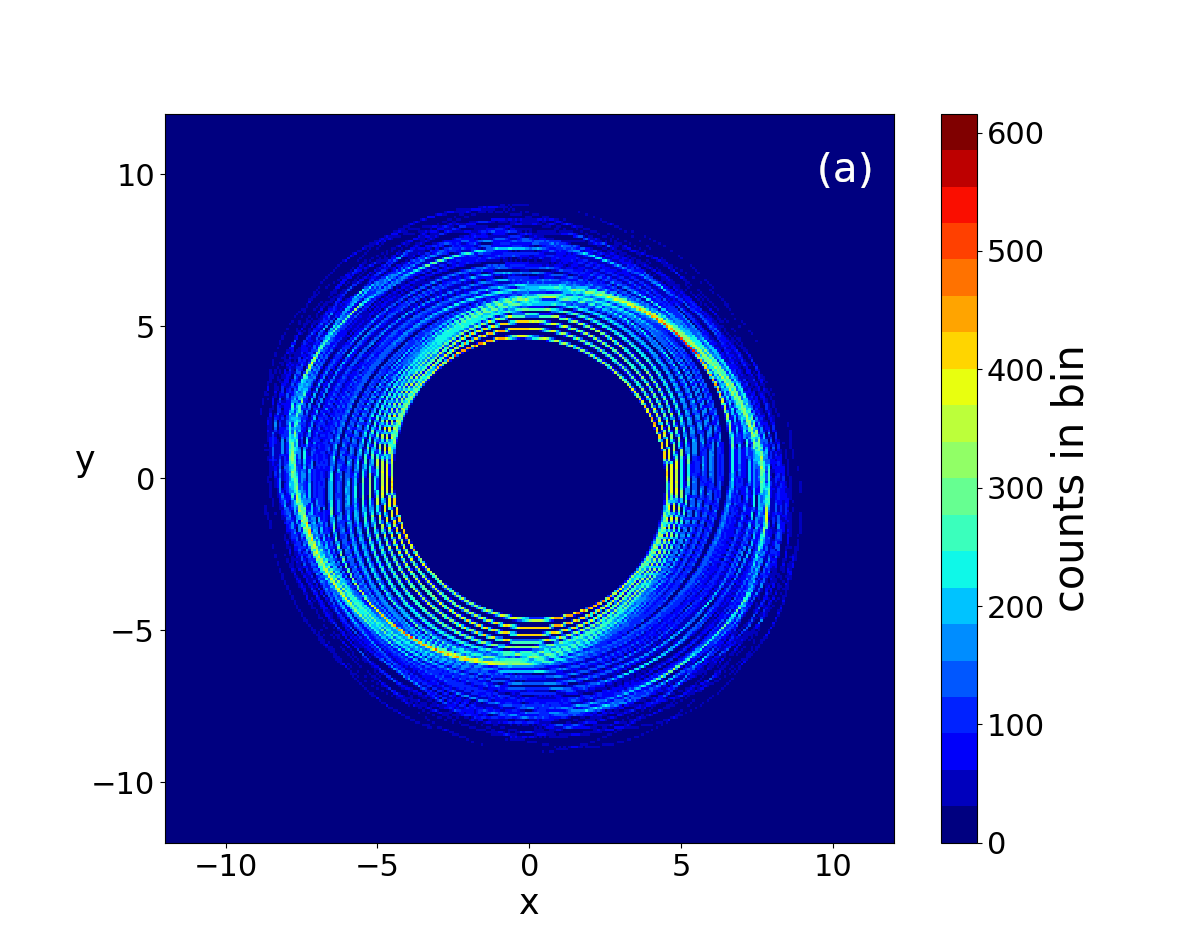}
\includegraphics[scale=0.30]{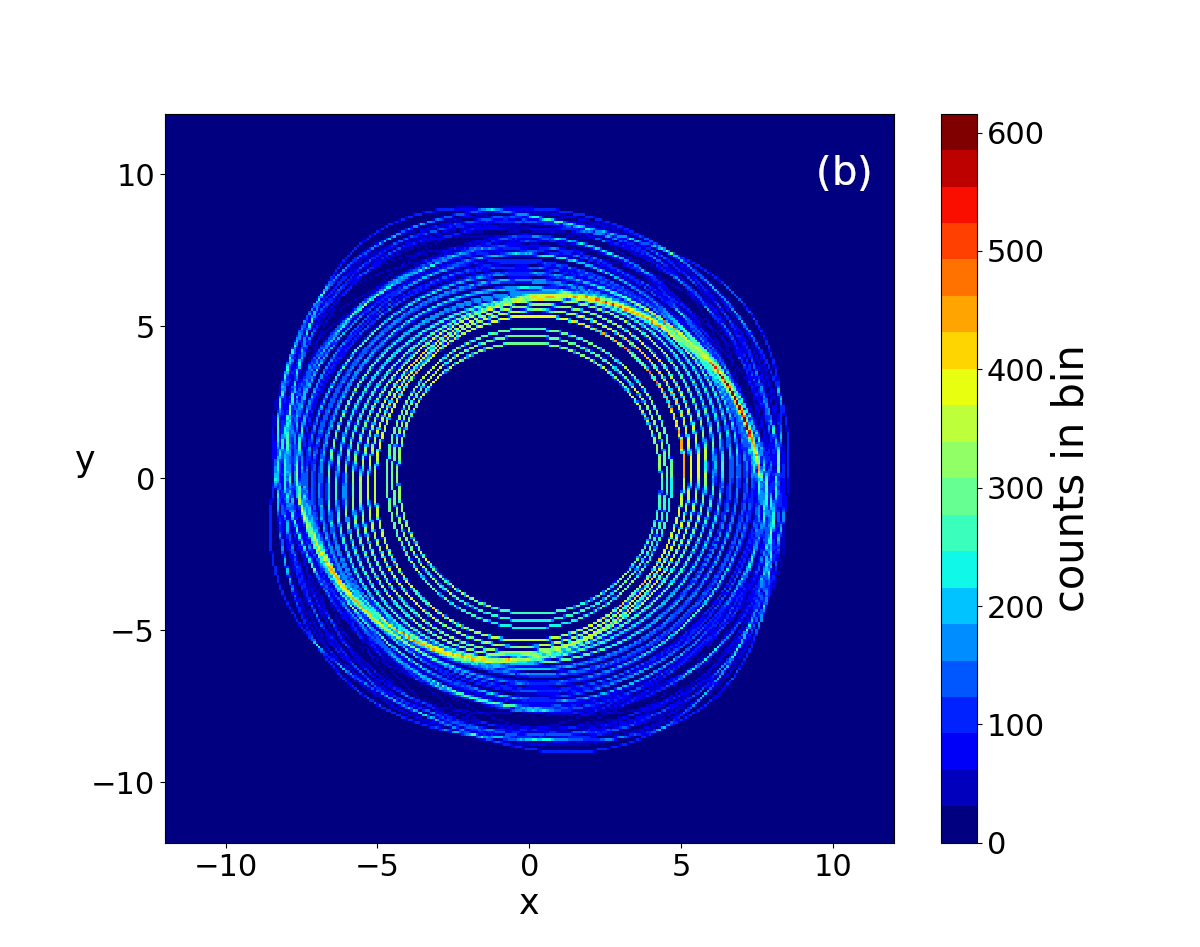}
\caption{Processed image of the spiral density wave for:  (a) $M_b=3 \times 
10^{10} M_\odot$, and (b) $M_b=1 \times 
10^{10} M_\odot$. In both cases $\Omega_{b}= 45$ ~km/s/kpc and $\Omega_{sp}=15$~km/s/kpc.}
\label{Fig5}
\end{figure*}

We plot the processed images of the spiral density waves, derived after the transformation to the old variables, for the values of the mass of the bar $M_b=3 \times 
10^{10} M_\odot$ in Fig. (\ref{Fig5})a  and  $M_b=1 \times 
10^{10} M_\odot$ in  Fig. (\ref{Fig5})b.
The main observation is that the  smaller the mass of the bar, the less intense the spiral density wave of the spiral arms, something that was expected since the coefficients $a_{m_1,m_2,m_3} (J_r)$ of the generating function $\chi$ in Eq. (\ref{gener}) depend on the mass $M_b$ of the bar. However, one important remark is that the spirals are well defined and present no breaks, as in the case of the Fig. \ref{Fig4}, with the greater value of the bar's mass $M_b$. It is now obvious that for even smaller values  than $M_b=1 \times 
10^{10} M_\odot$,  the spiral density wave will gradually disappear.

\subsection{The role of the pattern speed of the bar}
\begin{figure*}
\centering
\includegraphics[scale=0.29]{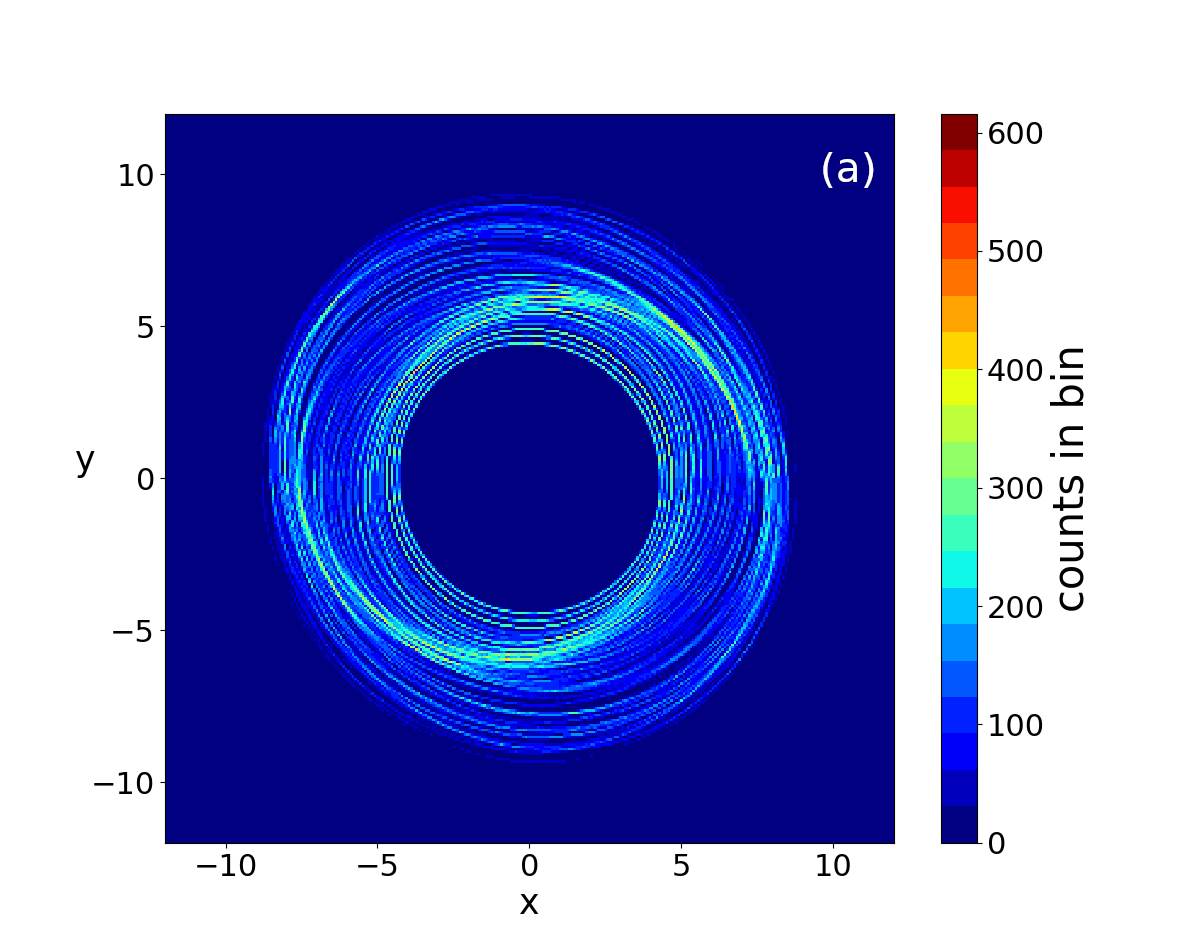}
\includegraphics[scale=0.21]{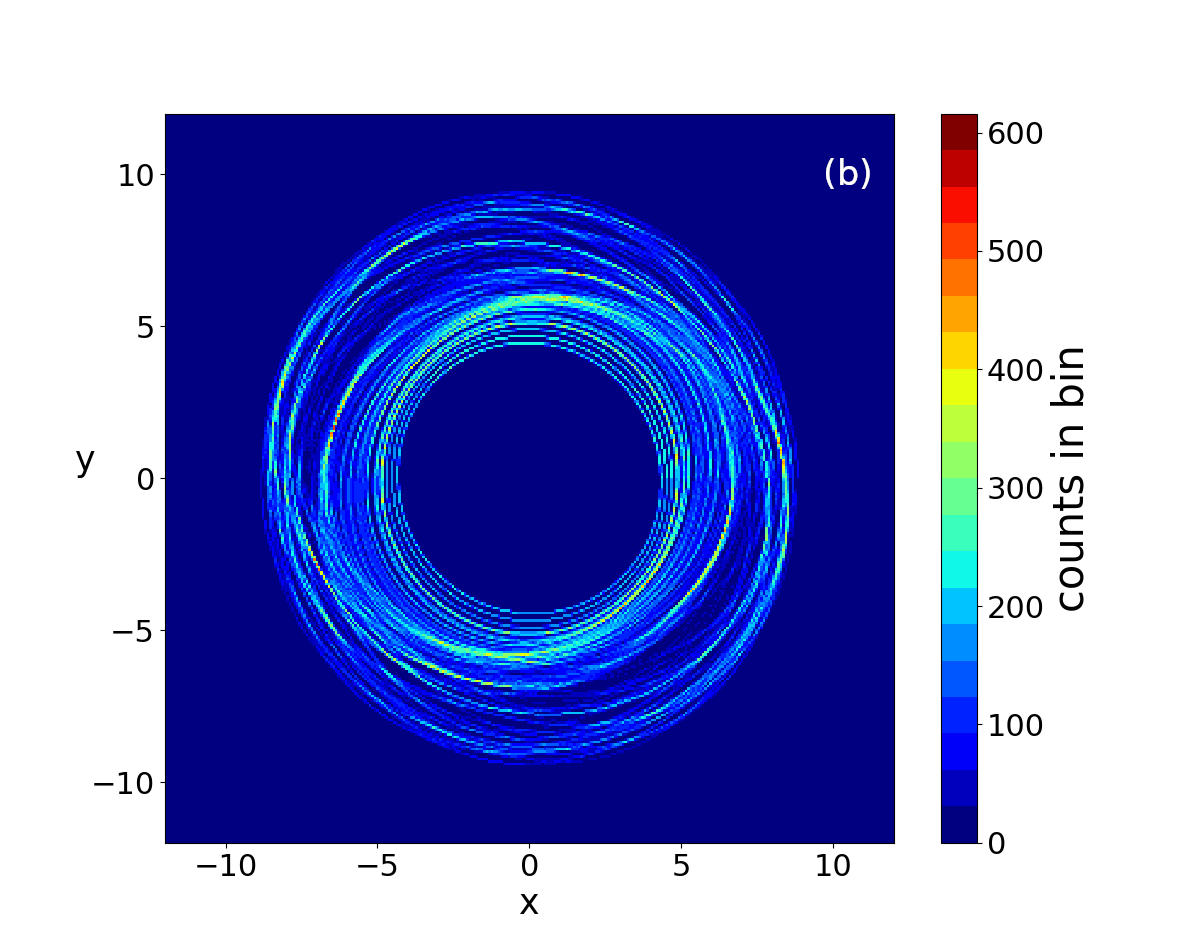}
\caption{ The spiral density wave derived from the periodic orbits of the $x_1$ family for the normal form Hamiltonian (\ref{hnew}) of  first order, for: (a) a  pattern speed of the bar $\Omega_b=20$~km/s/kpc, and  (b)a pattern speed of the bar $\Omega_b=60$~km/s/kpc. Both figures have $M_b=6.25 \times 
10^{10} M_\odot$ (a strong bar).} 
\label{Fig6}
\end{figure*}
In order to investigate the role of the pattern speed in the normal form construction of the time-dependent Hamiltonian (\ref{hamin}), we repeat the study of the previous subsection but for  different values of the pattern speed of the bar $\Omega_{b}$.
We now want to test a slower bar, namely  $\Omega_{b}$ =20~km/s/kpc, and a  faster one,  namely  $\Omega_{b}$ =60~km/s/kpc, keeping the bar's mass constant to $M_b=6.25 \times 
10^{10} M_\odot$ as in Fig. \ref{Fig4}. 
Observing in more detail Eq. (\ref{gener}) of the generating function $ \chi$, through which the transformation of the normal form is made, we see that the denominator includes the three basic frequencies, namely   $\Delta \Omega=\Omega_{sp}-\Omega_b$, $\kappa_c$, and $\Omega(r_c)-\Omega_{sp}$. If we reduce the absolute value of $\Delta \Omega=\Omega_{sp}-\Omega_b$ by using a smaller value of the pattern speed of the bar $\Omega_b$, the perturbation in the normal form construction is greater (a greater value of the generating function $\chi$) and this is reflected in the orbits having a greater deviation from the elliptical form. 

We construct the first-order normal form $H_{norm}^{(1)}$ of Eq. (\ref{hnew}) for  $\Omega_{b}$ =20~km/s/kpc (much slower bar rotation) and for $\Omega_{b}$ =60~km/s/kpc (faster bar rotation), and then we repeat the procedure described in Sect 3.2 in order to reveal the hidden structure  of the superposition of the deformed  precessing ellipses corresponding to the $x_1$ periodic families.  The spiral density waves derived are shown in Fig. \ref{Fig6}.  
  Fig. \ref{Fig6}a corresponds to $\Omega_b=20$~km/s/kpc, while Fig. \ref{Fig6}b correpsonds to $\Omega_b=60$~km/s/kpc. We observe that for the case of  the slower bar ($\Omega_b=20$~km/s/kpc), the spiral density wave derived is much less intense compared  with Fig. \ref{Fig4}, and less well defined as the spiral arms seem to be broken at several radii.  
  On the other hand, in Fig. \ref{Fig6}b, where the pattern speed of the bar has a much greater value ($\Omega_b=60$~km/s/kpc), the spiral density wave seems to be more intense 
but again it is much less well defined than in Fig. (\ref{Fig4}). So, there are some upper and lower limits in the value of the pattern speed of the bar in order to have an intense and well-defined spiral density wave structure in a galactic model with two different  pattern speeds.  
 All the calculations were performed on the "superpc" computer with ten CPU cores and 20 threads
at the Research Center for Astronomy and Applied Mathematics of the Academy of Athens.

\section{Conclusions}

In the present paper, we consider a galactic potential of a barred spiral galaxy with two different pattern speeds (for the bar and the spiral arms) and we eliminate the  dependence on time of the potential by considering the bar as a perturbation of the spiral potential, in the case where the bar rotates much faster that the spiral arms. The idea is to see how the precessing ellipses that support the spiral structure in the case of a grand design galaxy will be deformed and whether a  spiral density wave can still be detected. 
We use a Milky Way-like model with parameters chosen from \cite{Pett2014} and construct a normal form Hamiltonian in first- and second-order approximations, eliminating the time dependence due to the difference between the two pattern speeds. 

We find the stable periodic orbits of the $x_1$ family in this new normal form Hamiltonian and construct the spiral density wave made out of the approximately elliptical orbits of this family. Using the generating function $\chi$, we make a back transformation of these orbits to their original variables where the time is involved. The elliptical periodic orbits are deformed in elliptical rings having a certain thickness. However, if we superimpose all the orbits corresponding to several radii, and using  
 an image processing method, we reveal a slightly deformed spiral density wave that still survives. 
 This result  proves that in barred spiral galaxies where the bar rotates much faster that the spiral arms, the spiral structure can be supported by deformed precessing ellipses, such as in the case of grand design galaxies.
 
 We make a parametric study for the values of the bar's mass and pattern speed. By reducing the bar's mass, we find less intense but better defined spiral arms. A lower limit of the bar's mass is approximately the value $M_b=1 \times 
10^{10} M_\odot$. By reducing the bar's pattern speed,  we again find less intense spiral density waves and breaks in the spiral arms. Breaks in the spiral density wave appear also in the case of an extremely fast rotating bar (compared with the pattern speed of the spiral arms). 
The best, well-defined and intense, spiral density wave appears for a ratio of pattern speeds $3:1$ and a strong bar ($M_b=6.25 \times 
10^{10} M_\odot$).  

In the case where $\Omega_b=\Omega_{sp}$, the bar can no longer be considered as a perturbation of the spiral potential. This case has been already studied in the framework of the "manifold theory" in a series of papers, as mentioned in the introduction. According to this theory, in the case of barred spiral galaxies with one pattern speed, chaotic orbits with initial conditions along the unstable asymptotic manifolds, emanating from the Lagrandian points $L_1$ and $L_2$, as well as nearby sticky chaotic orbits, can support the spiral structures for a long time.

  This study is an alternative scenario to the manifold theory, where the spiral potential is considered as a perturbation of the bar potential (see \cite{Efth2020}) in the case of a barred spiral galactic model having two different pattern speeds. Here, in contrast,  we consider the bar potential as a perturbation of the spiral potential in the case where the bar rotates fast enough, compared with the pattern speed of the spiral arms.  This is a proposed scenario for cases of barred spiral galaxies where the spiral arms have a well-defined symmetric shape, such as the one proposed for the Milky Way.
  In \cite{ger2011}, which is a review of the pattern speeds of the Milky Way, the author concludes that the bar of the Milky way is a fast rotator having  between double and triple the speed of the spiral arms. In our study we adopt values for the pattern speeds of the bar and the spiral arms that are suggested in Gerhard's review paper. In the cases where the bar rotates much faster than the spiral arms, as in the case of the Milky Way, the arms in general seem more symmetrical and well defined, having almost the same pitch angle along the entire radius of the galaxy (see also \cite{Font2019} where they give the pattern speeds of the bar and the spiral arms of  a fairly large sample of barred spiral galaxies). This image of spiral arms can be supported by ordered elliptical orbits. This is a scenario that is consistent with the density wave theory. \cite{Font2019} also found that the longer the bars (and as a consequence slower), the smallest the differences between the pattern speeds. Thus, the spiral arms, which are most likely bar driven, seem less symmetric with variable pitch angle, while some other features appear such as gaps, bridges, and bifurcations. This image is more consistent with the scenario of chaotic orbits along unstable manifolds  supporting the spiral structures. Further numerical investigation is needed in order to confirm such a statement.
    
     So, in conclusion, for the case where the bar rotates much faster than the spiral arms, we suggest an alternative scenario to the "manifold theory" for supporting the spiral structures of the galaxy, which is, in fact, a generalization of the density wave theory.

\end{document}